\documentclass[%
    reprint,
    superscriptaddress,
    longbibliography,
    nofootinbib,
    amsmath,
    amssymb,
    aps,
    floatfix,
]{revtex4-2}

\usepackage{graphicx}
\usepackage{dcolumn}
\usepackage{bm}
\usepackage[utf8]{inputenc}
\usepackage{acronym}
\usepackage{graphicx}
\usepackage{amsmath,amsfonts,amssymb}
\usepackage{braket}
\usepackage{tikz}
\usetikzlibrary{quantikz}
\usepackage{acronym}
\usepackage{xspace}
\usepackage{siunitx}
\usepackage{booktabs}
\usepackage[hidelinks]{hyperref}
\usepackage{nicefrac}       
\usepackage{microtype}      
\usepackage{mathtools}
\usepackage{listings}
\usepackage{todonotes} 
\usepackage{siunitx}
\usepackage{multirow}
\usepackage{tabularx}

\newcommand{\GeV}{\ensuremath{\,\text{Ge\hspace{-.08em}V}}\xspace}

\newcommand{\PYTHIA} {{\textsc{pythia}}\xspace}

\newcommand{\FASTJET} {{\textsc{FastJet}}\xspace}

\newcommand{\pt}{\ensuremath{p_{\mathrm{T}}}\xspace}
\newcommand{\kt}{\ensuremath{k_{\mathrm{T}}}\xspace}
\newcommand{\fbinv} {\mbox{\ensuremath{\,\text{fb}^{-1}}}\xspace}
\newcommand{\mjj}{\ensuremath{m_\mathrm{jj}}\xspace}

\newacro{LHC}{Large Hadron Collider}
\newacro{HEP}{High Energy Physics}
\newacro{QML}{Quantum Machine Learning}
\newacro{ML}{Machine Learning}
\newacro{SM}{Standard Model}
\newacro{BSM}{Beyond Standard Model}
\newacro{MC}{Monte Carlo}
\newacro{AD}{Anomaly Detection}
\newacro{RS}{Randall-Sundrum}
\newacro{AE}{autoencoder}
\newacro{QSVM}{Quantum Support Vector Machines}
\newacro{NISQ}{Noisy Intermediate Scale Quantum}
\newacro{TPR}{True Positive Rate}
\newacro{FPR}{False Positive Rate}

\newcommand{\congto}{\xrightarrow{\raisebox{-0.4ex}[0ex][0ex]{\tiny{$O$}}}}

\begin{document}

\title{Quantum anomaly detection in the latent space of proton collision events at the LHC}

\author{Vasilis Belis$^{*\dagger}$}
\affiliation{Institute for Particle Physics and Astrophysics, ETH Zurich, 8093 Zurich, Switzerland}
\thanks{Corresponding author: \href{mailto:vbelis@phys.ethz.ch}{vbelis@phys.ethz.ch}}
\author{Kinga Anna Wo\'zniak$^\dagger$}
\affiliation{European Organization for Nuclear Research (CERN), CH-1211 Geneva, Switzerland}
\affiliation{Faculty of Computer Science, University of Vienna, Vienna, Austria}
\altaffiliation{These authors contributed equally.}
\author{Ema Puljak$^\dagger$}
\affiliation{European Organization for Nuclear Research (CERN), CH-1211 Geneva, Switzerland}
\affiliation{Departamento de Física, Universitat Autònoma de Barcelona, 08193 Bellaterra (Barcelona), Spain}
\altaffiliation{These authors contributed equally.}
\author{Panagiotis Barkoutsos}
\affiliation{IBM Quantum, IBM Research – Zurich, 8803 R\"uschlikon, Switzerland}
\author{Günther Dissertori}
\affiliation{Institute for Particle Physics and Astrophysics, ETH Zurich, 8093 Zurich, Switzerland}
\author{Michele Grossi}
\affiliation{European Organization for Nuclear Research (CERN), CH-1211 Geneva, Switzerland}
\author{Maurizio Pierini}
\affiliation{European Organization for Nuclear Research (CERN), CH-1211 Geneva, Switzerland}
\author{Florentin Reiter}
\affiliation{Institute for Quantum Electronics, ETH Zurich, 8093 Zurich, Switzerland.}
\author{Ivano Tavernelli}
\affiliation{IBM Quantum, IBM Research – Zurich, 8803 R\"uschlikon, Switzerland}
\author{Sofia Vallecorsa}
\affiliation{European Organization for Nuclear Research (CERN), CH-1211 Geneva, Switzerland}

\date{\today}

\begin{abstract}
The ongoing quest to discover new phenomena at the LHC necessitates the continuous development of algorithms and technologies. Established approaches like machine learning, along with emerging technologies such as quantum computing show promise in the enhancement of experimental capabilities. In this work, we propose a strategy for anomaly detection tasks at the LHC based on unsupervised quantum machine learning, and demonstrate its effectiveness in identifying new phenomena. The designed quantum models—an unsupervised kernel machine and two clustering algorithms—are trained to detect new-physics events using a latent representation of LHC data, generated by an autoencoder designed to accommodate current quantum hardware limitations on problem size. For kernel-based anomaly detection, we implement an instance of the model on a quantum computer, and we identify a regime where it significantly outperforms its classical counterparts. We show that the observed performance enhancement is related to the quantum resources utilised by the model.

\end{abstract}

\maketitle

\section{Introduction}\label{sec:intro}

\ac{QML} is a nascent field at the intersection of quantum information processing and machine learning~\cite{Rebentrost2014, Biamonte2017,Schuld_2018_book, Schuld_2019, circuit_centric_Schuld2020, Havlicek2019, Lloyd_2020}. It has the potential to revolutionise the way we solve problems in \ac{HEP}, where the particle collision data is generated by fundamentally quantum processes; the particle interactions. There is a growing number of studies investigating \ac{QML}, its possible advantages over classical computing, and its applicability to varied real-life problems~\cite{peters2021, Huang2021, Liu2021rigorous, Muser:2023tos, huangQA2022, Kubler_2021, caro_generalization_2022, cong_quantum_2019, dimeglio2023quantum}.

The use of quantum computing techniques on \ac{HEP} problems was introduced by Ref.~\cite{Mott:2017xdb}, where the task of training a classifier for photon selection in a search for the Higgs boson was framed as a quantum annealing problem. Since then, \ac{QML} algorithms have been examined for event reconstruction tasks~\cite{Tuysuz_2020, Tuysuz_2021, magano_quantum_2022, Lerjarza2022, duckett2022} and classification problems~\cite{Guan_2021, wu_2021_kernel, terashi_event_2021, blance_quantum_2021, qmlHiggs2021} in \ac{HEP}. All these problems belong to the realm of supervised learning, in which the training process is guided by comparing the algorithm output to the ground truth, provided with the training dataset. A natural evolution of this line of research is to investigate QML techniques for unsupervised problems, i.e., learning tasks on unlabeled data. 

\begin{figure*}[t]
    \centering
    \includegraphics[width=0.9\textwidth]{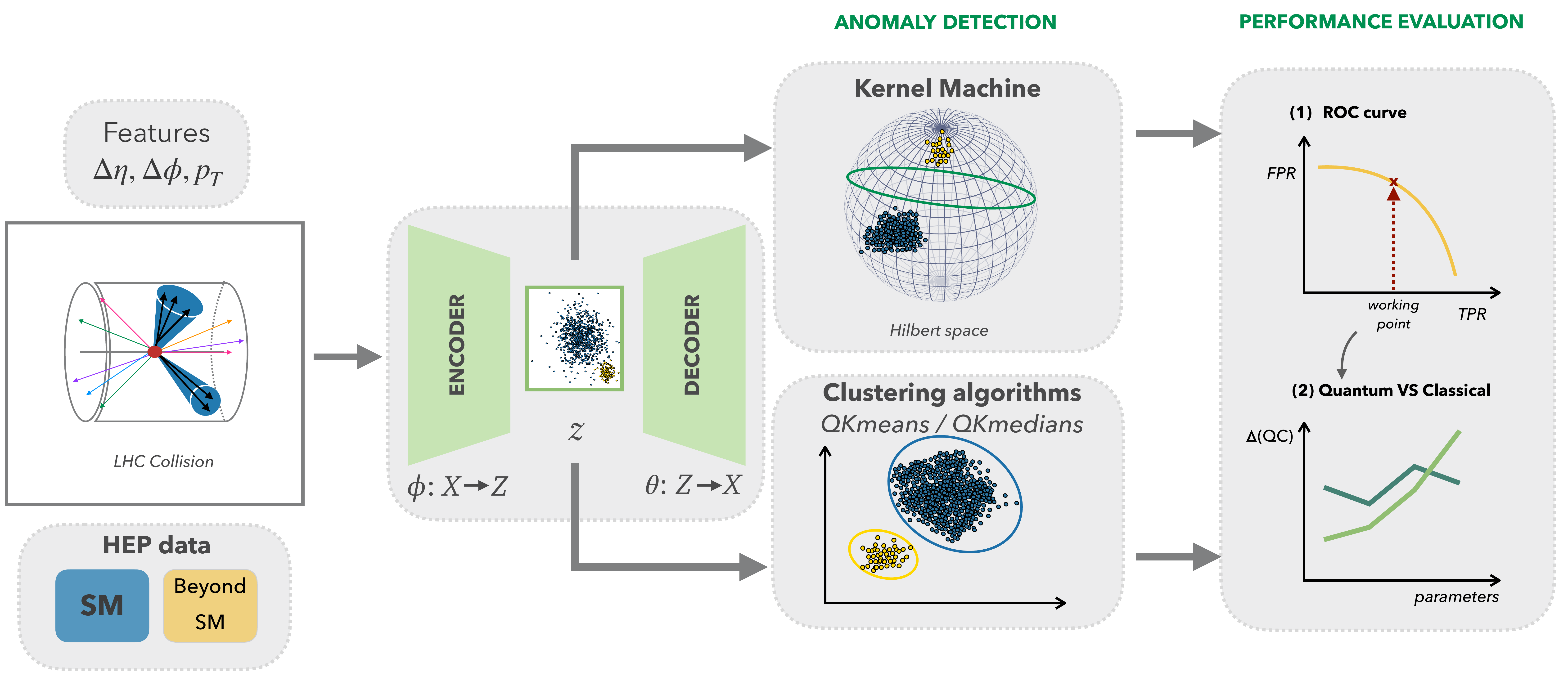}
    \caption{\textit{Classical-quantum pipeline.} LHC collision data (simulation) are passed through an autoencoder for dimensionality reduction followed by the quantum anomaly detection models: \textit{unsupervised quantum kernel machine} and \textit{quantum clustering algorithms} (QK-means/QK-medians). Each jet contains 100 particles, each particle is described by three features $(\Delta\eta, \Delta\phi, p_T)$ where $\Delta$ represents a distance from the jet axis. Hence, a dijet collision event is described by 300 features. The quantum models are trained on Standard Model data and learn to recognise anomalies in unseen data. All models are evaluated by calculating the Receiver Operating Characteristic (ROC) curve and metrics appropriate for anomaly detection tasks, and are compared to their classical counterparts (see ``Evaluation of model performance" subsection in the Results).} 
    \label{fig:workflow}
\end{figure*}

In this work, we investigate the use of a \textit{classical-quantum pipeline}, in which a high-dimensional dataset is projected to a latent space via a classical autoencoder and a set of unsupervised quantum algorithms are trained on the latent-space features. We consider the problem of searching for new physics processes in \ac{LHC} events as an anomaly detection problem. Our approach aims to implement a prototype pipeline that is able to process realistic datasets and could be eventually deployed at the \ac{LHC} to assist scientific discovery (see Fig.~\ref{fig:workflow}). Similar problems have been considered on simpler low-dimensional \ac{HEP} datasets where different approaches have been followed~\cite{Blance_ad_2021, alvi_quantum_2022, Ngairangbam_2022}.

After the discovery of the Higgs boson by the ATLAS~\cite{atlas_higgs, higgs_anniversary_nat_atlas} and CMS~\cite{cms_higgs, higgs_anniversary_nat_cms} collaborations, one of the main goals of the \ac{LHC} physics program is the search for new phenomena that would answer some of the open questions associated with the \ac{SM} of particle physics, the theory that describes our understanding of the fundamental constituents of the universe: why are the known forces in nature characterised by largely different energy scales? What is the mechanism that generates neutrino masses? What is the origin of Dark Matter and Dark Energy? 
To answer these questions, proton beams are accelerated in the \ac{LHC} up to an energy of 6.75 TeV and made to collide. These collisions happen at energies that were typical of particle collisions in the early universe. By studying them in a controlled environment, physicists aim to understand post-big-bang physics and possibly highlight the difference between observations and SM predictions. 
These searches for \ac{BSM} physics are typically \emph{model-dependent}. In this context, a \emph{signal} is defined as a specific process described by the chosen \ac{BSM} scenario. The \emph{background} is defined as any \ac{SM} process that generates a similar detector signature as that of a \ac{BSM} process. In the recent past, the use of supervised \ac{ML} algorithms has become prominent. These \ac{ML} algorithms are typically trained on labeled data from \ac{MC} simulations and applied to experimental data. 

This model-dependent approach exploits at best the solid understanding of the physics associated with the postulated signal and the known background processes, and typically reaches remarkable signal-to-background discrimination power, as in the case of the discovery of the Higgs boson~\cite{cms_higgs, atlas_higgs}. However, the need to postulate a priori the signal of interest has an intrinsic drawback: searches based on a given signal assumption are typically less sensitive to other kinds of signals. If an unexpected signature is present in the data, it might not be identified due to the inherent bias of the study toward the chosen signal hypothesis.

To overcome this drawback, an unsupervised approach can be adopted, where the search for new physics is reframed as an \ac{AD} task. This way, \ac{BSM} searches can be constructed to rely minimally on specific new physics scenarios. In this context, \emph{anomalies} are defined as events that collectively signify the presence of rare \ac{BSM} phenomena and have an extremely low probability under the hypothesis that only particles and interactions predicted by \ac{SM} are present in the data.  
Anomaly detection for new physics searches in multijet events has been proposed in Refs.~\cite{collins2018,DAgnolo_Wulzer_2019,Heimel:2018mkt,Farina_2020} and then refined in many other studies~\cite{Kasieczka_2021,dark_machines2022}. This strategy has been introduced as a new way to select events in real time~\cite{Cerri_2019} and it has been successfully adapted in a proof-of-concept study with real data~\cite{Knapp_2021}. Recently, the ATLAS collaboration released two searches with weakly supervised~\cite{ATLAS:2020iwa} and unsupervised~\cite{ATLAS:2023azi, ATLAS_ad_data2023} techniques.

Unsupervised algorithms for \ac{AD} rely on less information than supervised methods since signal vs. background characterisation through labels is not provided. On the other hand, thanks to their increased generalisability to other BSM signatures, \ac{AD} techniques offer qualitative advantages with respect to traditional methods, which can contribute to expand the physics reach of the \ac{LHC} experiments~\cite{Kasieczka_2021, dark_machines2022,Belis:2023mqs}. This is why \ac{AD} algorithms for HEP offer an interesting scientific problem to investigate, particularly with unsupervised QML techniques. 
Quantum algorithms have many specific aspects that differentiate them from classical solutions proposed in the literature, typically based on deep neural networks trained on large datasets. In particular, QML provides advantages in terms of computational complexity~\cite{lloyd2013quantum, Liu2021rigorous}, generalisation with few training instances~\cite{caro_generalization_2022}, or the ability to uncover patterns unrecognisable to classical approaches~\cite{huangQA2022, Gao_2022}. 

Similar to other prototype studies of \ac{AD} techniques in HEP~\cite{Kasieczka_2021}, we consider the problem of looking for new exotic particles decaying to jet pairs. Jets are sprays of close-by particles originating from a shower of hadrons, following the production of quarks and gluons in \ac{LHC} collisions. Traditional jets have a one-prong cone shape and are copiously produced in so-called Quantum Chromodynamics (QCD) multijet events, the most frequent processes occurring in \ac{LHC} collisions. At the \ac{LHC}, multi-prong jets could emerge from all-quark decays of heavy particles manifesting as a peak in the dijet-mass ($m_{jj}$) distribution (a {\it resonance}). However, this peak could be obscured by the huge multijet background and an \ac{AD} technique could be used to suppress the background and make the resonant peak emerge. 

In their searches for new-physics signatures, LHC experiments routinely employ models that compress the data~\cite{ATLAS_ad_data2023, ATLAS:2023azi,CMS-PAS-EXO-22-026, Farina_2020, Roy:2019jae, Heimel:2018mkt, Knapp_2021, Cheng2023, Blance:2019, Bortolato:2021}. This involves using autoencoder models, where the discrepancy between the encoder and decoder networks serves as a measure to detect anomalies. Such models have been also proposed for the real-time event selection system (trigger) of the LHC experiments~\cite{Govorkova_2022, Cerri_2019}.

In our study, we employ quantum anomaly detection algorithms to identify anomalies in the latent space generated by the encoder network of our autoencoder model (see Fig.~\ref{fig:workflow}). We develop an \textit{unsupervised quantum kernel machine} and two \textit{quantum clustering} algorithms to identify latent representations of \ac{BSM} processes in a model-independent setting. The discrimination power of the various models is assessed using a metric tailored to anomaly detection tasks. 
In the case of kernel-based anomaly detection, we find that the ability of our quantum model to identify new-physics events is significantly enhanced as more quantum resources are utilised. 
Furthermore, we note that as we increase the amount of entanglement and the number of qubits in the designed data encoding circuit beyond a certain threshold, the quantum model surpasses its classical counterpart in performance.
The statistical robustness of our results is systematically studied in numerical experiments across the parameters of interest.

\section{Results}
\subsection{Data}\label{sec:data-section}
The dataset used for this study consists of proton-proton collisions simulated at a center-of-mass energy $\sqrt{s}=13$~TeV (see also Methods). The following BSM processes are representative of new-physics scenarios that the \ac{LHC} experiments are sensitive to. We consider them  as anomalies to benchmark the model performance.
\begin{itemize}
    \item Production of narrow Randall-Sundrum gravitons~\cite{Randall:1999ee} decaying to two $W$-bosons (Narrow $G\to WW$).
    \item Production of broad Randall-Sundrum gravitons~\cite{Randall:1999ee} decaying to two $W$-bosons (Broad $G\to WW$).
    \item Production of a scalar boson $A$ decaying to a Higgs and a Z bosons ($A\to HZ$). Higgs bosons are then forced to decay to $ZZ$, resulting in a $ZZZ$ final state.
\end{itemize}

All the $W$ and $Z$ bosons are forced to decay to quark pairs, resulting in all-jet final states. \ac{SM} events are generated emulating QCD multijet production at the \ac{LHC}, by far the most abundant process in a sample of all-jet events. Detector coordinates are specified with respect to the Cartesian coordinate system with the $z$-axis oriented along the beam axis and $x$ and $y$ axes defining the transverse plane. 
The azimuthal angle $\phi$ is computed from the $x$-axis. 
The polar angle $\theta$, measured from the positive $z$-axis, is used to compute the pseudorapidity $\eta = -\log(\tan(\theta/2))$. The transverse momentum $\pt$ is defined as the projection of the momentum of a particle or a jet on the plane perpendicular to the beam axis. For more details regarding the construction of the dataset see the Methods section.

\subparagraph{\textit{Dimensionality reduction}.}
\label{sec:DimRed}
Direct loading and processing of such datasets on current near-term, noisy, quantum devices is typically not possible due to the limited number of qubits and their small decoherence times~\cite{Preskill_2018}. To address this challenge, we develop a convolutional \ac{AE} that maps events into a latent representation of reduced dimensionality (Methods). The \ac{AE} is designed to act at \textit{per-particle} level in its 1D convolutional input layer, to be appropriately biased to the structure of our jet data. For our analysis, the reduced dijet datasets have $2\ell$ features, where $\ell$ is the dimension of the latent representation for each jet.

Most studies of \ac{QML} applications in \ac{HEP} and beyond, rely on pre-processing methods to reduce the dimensionality of features. Even though such a task is presently required for a high-dimensional dataset, it can introduce biases that negatively affect the performance of the \ac{QML} models. Many studies report similar performance between quantum and classical algorithms~\cite{Guan_2021, qmlHiggs2021, blance_quantum_2021, terashi_event_2021, wu_2021_kernel, alvi_quantum_2022, Ngairangbam_2022}. We expect that quantum models require access to the unprocessed data, ideally right after measurement by the detector, to exploit correlations and potentially achieve higher performance.
The more post-processing steps there are after the measurement, the higher the probability of losing fundamental quantum correlations leading to suboptimal performance of the quantum algorithms~\cite{Belis:2024ctj}. For this reason, we work exclusively with low-level data, that is, the four-momenta of the particles. The autoencoder is applied to the input data, which is still a form of classical post-processing. However, unlike PCA or manual feature selection, autoencoders are able to retain, at least partially, the non-linear correlations in the latent space. 

\subsection{Models for quantum anomaly detection}\label{sec:methods-ad}

The developed \ac{QML} algorithms are trained to define a metric of typicality for QCD jets, which can then be used to identify \ac{BSM} anomalies.
We consider two categories of quantum models: \textit{kernel machines} and \textit{clustering algorithms}.

\subparagraph{\textit{Quantum Kernel machines}.}Support Vector Machines (SVM) are powerful models that can serve as the kernel-based model blueprint in supervised learning~\cite{svmVapnik}. \ac{QSVM} are defined by constructing a \textit{quantum kernel}~\cite{Schuld_2019, Havlicek2019}, and have been recently employed for classification tasks in \ac{HEP}~\cite{wu_2021_kernel, qmlHiggs2021}. We extend this idea to the unsupervised learning setting, where we separate expected (\ac{SM}) and anomalous (\ac{BSM}) events (cf. Methods). 

\begin{figure}[htb]
\centering
\includegraphics[width=\linewidth]{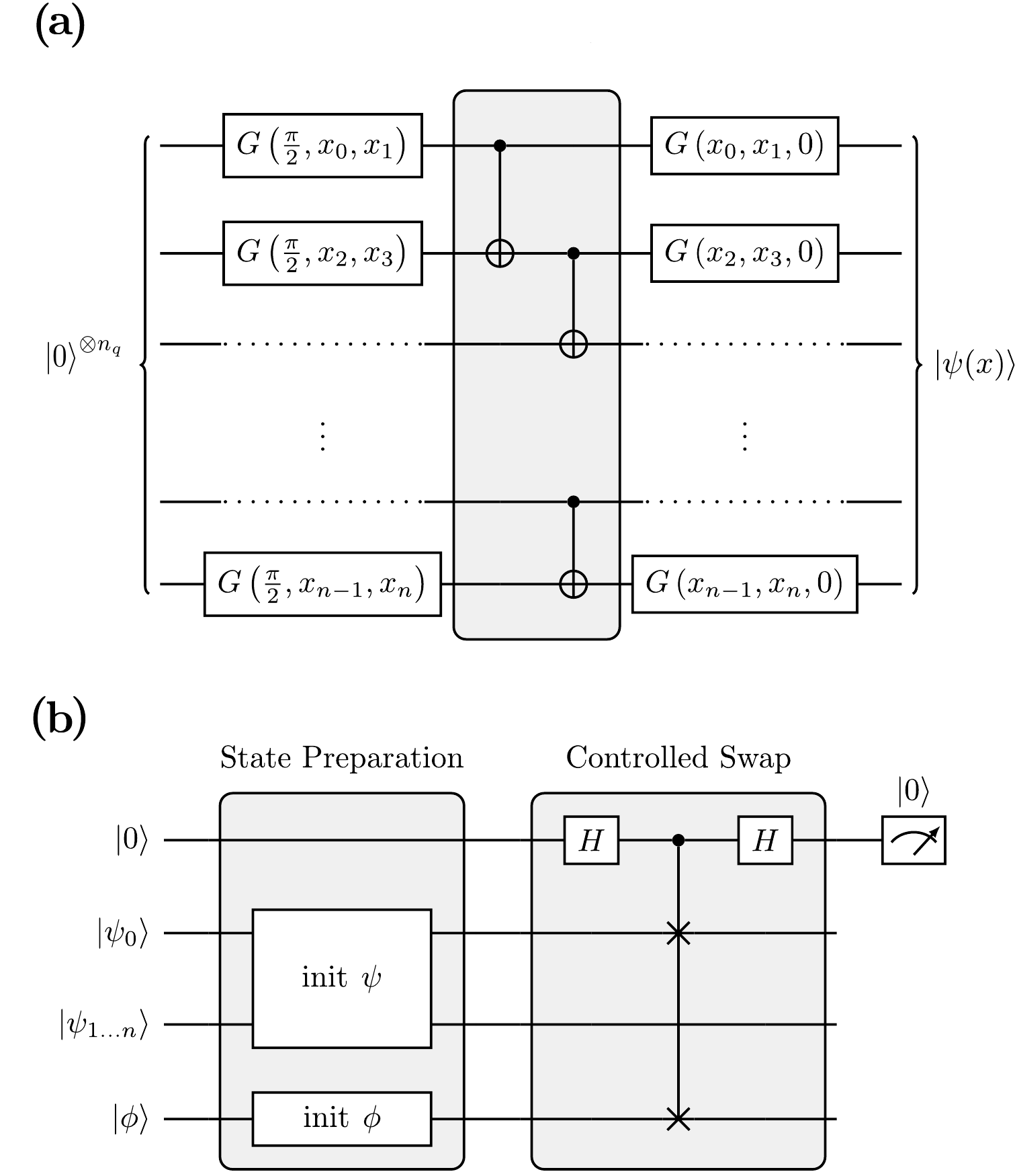}
                \caption{\textit{The quantum circuits.} (\textbf{a}) Data encoding circuit $U(x)$, for a data point $x$, that implements the feature map of the unsupervised kernel machine and is used to define the quantum kernel $k(x_i,x_j)
= 
\left|{\braket{0|U^\dagger(x_i) U(x_j)|0}}\right|^2$, where $G(\theta,\phi,\lambda)\in\text{SU(2)}$ is a universal 1-qubit gate, and $x_i$, for $i=0,1,\dots, n$, denotes the elements of the input feature vector $x$.  The entanglement gates correspond to CNOT gates. (\textbf{b}) Quantum distance calculation circuit used to compute the similarity between an input sample and a cluster center in the QK-means algorithm. The prepared $\ket{\psi}$ and $\ket{\phi}$ states depend on the input feature vectors (Methods).}
                \label{fig:circuits}
                \end{figure}

Different strategies to embed data into quantum states have been studied in literature~\cite{Havlicek2019,dataEncoding2020,Lloyd_2020}. We design a quantum feature map implemented by the quantum circuit depicted in Fig.~\ref{fig:circuits}a, taking into account the following properties. The data embedding circuit encodes two of $n$ input features in the two physical degrees of freedom of each qubit via unitary rotation gates and employs nearest-neighbors entanglement between qubits. The encoding gates are repeated, by permuting the rotation angles to which the features correspond, introducing non-trivial interactions between features, and avoiding encoding different features along the same axis of rotation. The quantum circuit, as a whole, can be repeated $L$ to increase the \textit{expressibility} and \textit{entanglement capability} of the feature map; for more details see Methods section, the Supplementary Note 1, and Ref.~\cite{sim_expressibility_2019}. 
The number of qubits $n_q$ together with the repetitions $L$ of the data encoding circuit quantify the amount of quantum resources that are utilised by the model, as further discussed in the ``Evaluation of model performance" subsection.  
Furthermore, the quantum circuit is designed to be hardware-efficient, i.e., it is compact and takes into account the topology of superconducting qubits.

\subparagraph{\textit{Clustering algorithms}.} The Quantum \textit{K-means} (QK-means) and Quantum \textit{K-medians} (QK-medians) are unsupervised clustering algorithms~\cite{qkmeans_noisy}. In resemblance to their classical counterparts, these algorithms partition a dataset of $N_{\mathrm{train}}$ samples $x \in \mathbb{R}^n$ into $k$ clusters based on a distance metric. For both clustering models, we perform \emph{amplitude encoding} to map the inputs to normalised quantum states, requiring $n_q=\log_2(2n)$ qubits for a dataset of $n$ features. We follow two different strategies for the cluster assignment.

In the case of the QK-means algorithm, each cluster center is defined as the mean of the training samples assigned to it in the previous iteration. For the cluster assignment we first use the distance calculation quantum circuit presented in Fig.~\ref{fig:circuits}b and perform optimisation with a quantum algorithm~\cite{durr1999quantum, boyer1998tight}, based on Grover's search~\cite{grover1996fast}, to find the closest cluster to each training sample. For the distance calculation in the case of \textit{Quantum K-medians} (QK-medians), we choose a shallower quantum circuit with state preparation part and only one Hadamard gate~\cite{qkmeans_noisy} and a hybrid quantum-classical minimisation procedure consisting of classical iterative algorithm for finding the median~\cite{l1_median_real}. For more details see the corresponding Methods section.

\subsection{Evaluation of model performance}\label{sec:results}
\begin{figure*}[th]
    \centering
    \includegraphics[width=1.01\textwidth]{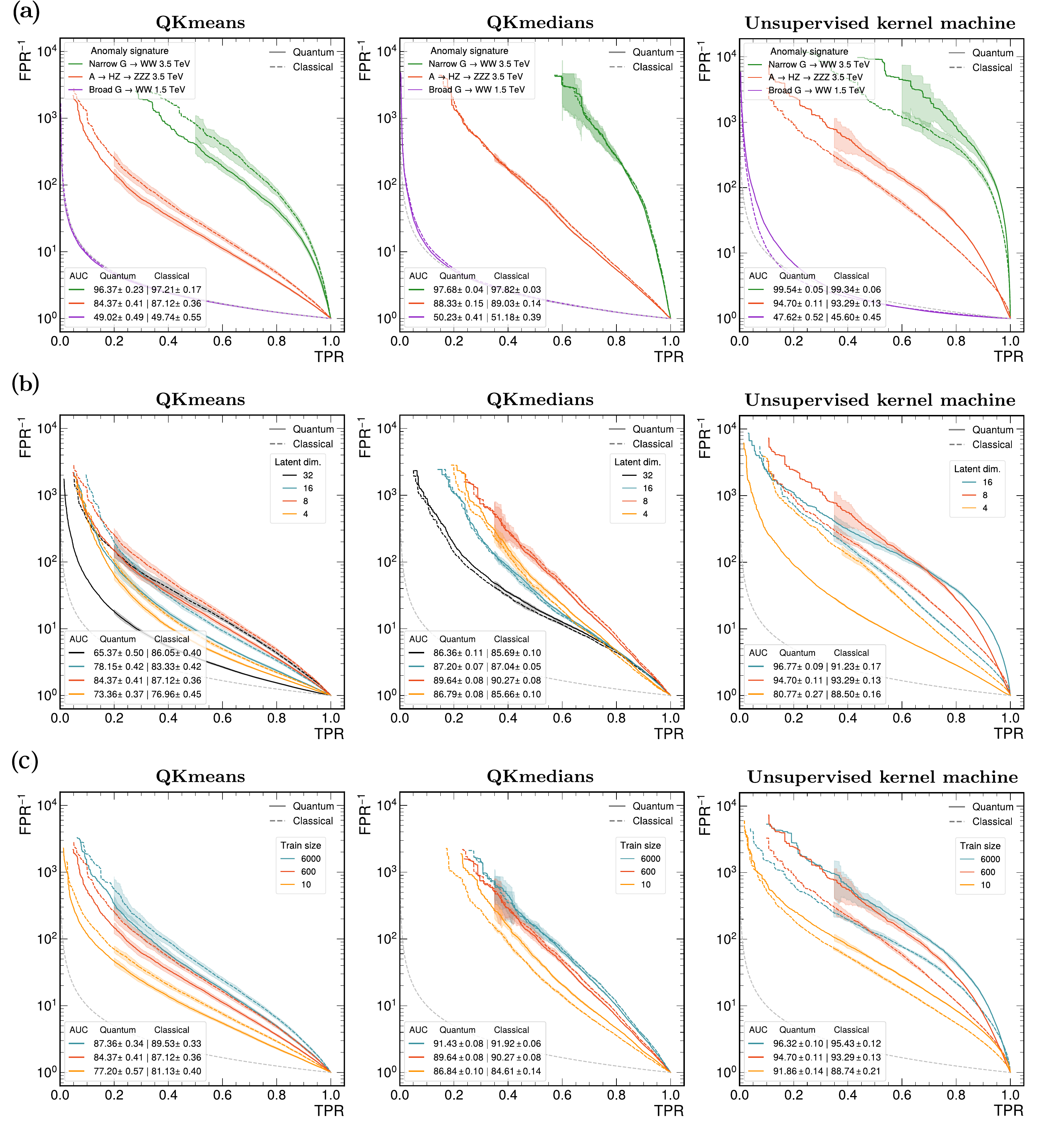}
    \caption{\textit{Performance evaluation results.} Each subplot displays the Receiver Operating Characteristic (ROC) curve and the corresponding Area Under the Curve (AUC) on test data for each model and each parameter of interest. Rows represent the model evaluation as a function of the parameters of interest: (\textbf{a}) new-physics anomalous signatures, (\textbf{b}) latent space dimension, and (\textbf{c}) the number of training samples. Columns correspond to the different anomaly detection models. A 5-fold testing is performed to assess the statistical significance of the results and of the differences in performance, using a test dataset of $10^5$ samples where half of the samples are anomalies and half are \ac{SM} events. The best-performing classical model is a kernel model equipped with the Radial Basis Function (RBF) kernel. The uncertainty bands represent one standard deviation and are drawn only for the True Positive Rate (TPR) range of interest. For smaller values of TPR, the uncertainties increase due to low testing statistics and the bands are omitted for readability purposes.}
    \label{fig:results}
\end{figure*}

\begin{figure*}[thb]
\centering
    \includegraphics[width=\linewidth]{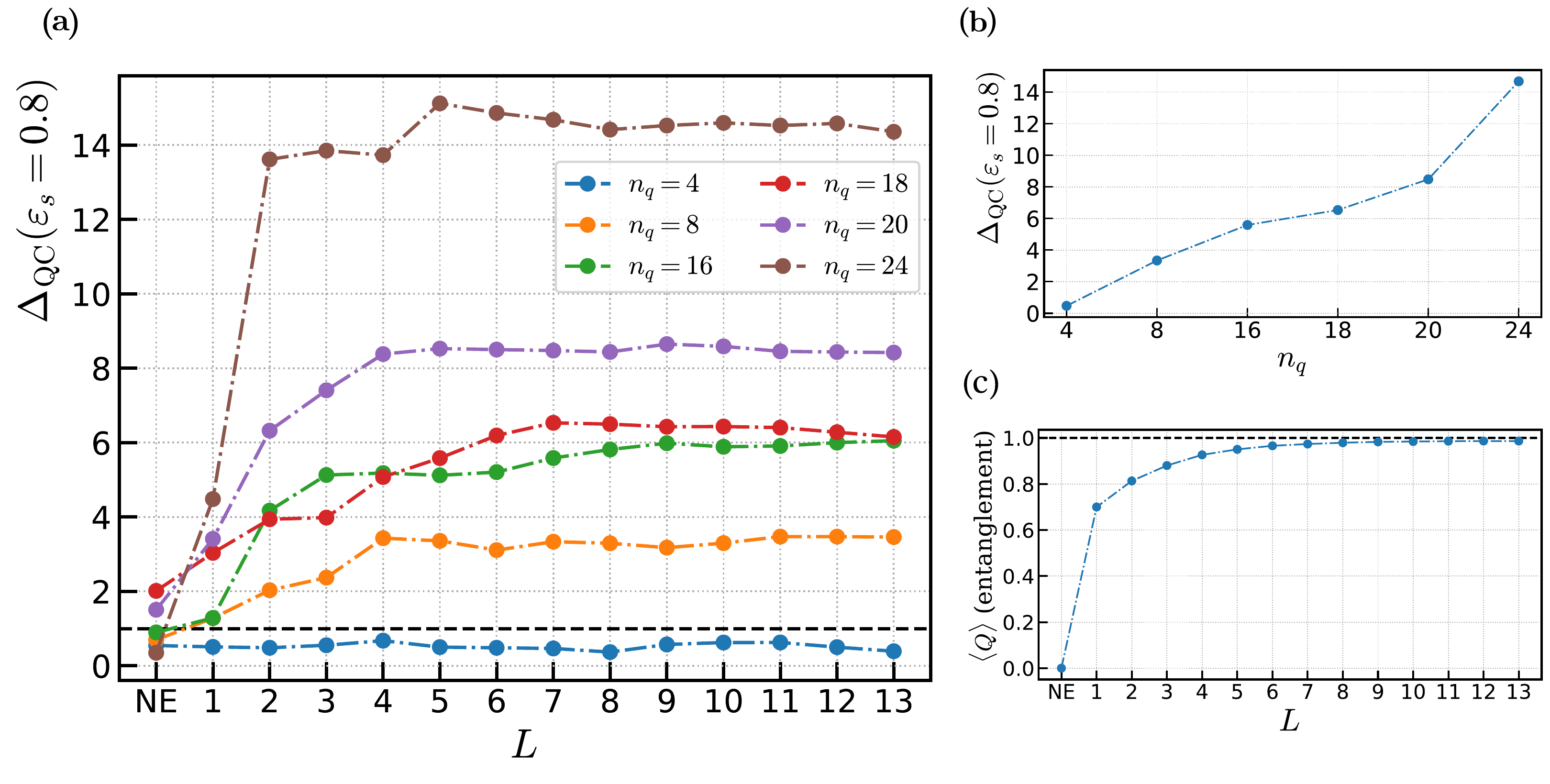}
    \caption{\textit{Performance of the unsupervised quantum kernel machine and role of entanglement}. (\textbf{a}) The performance of the unsupervised quantum kernel machine, quantified by $\Delta_\mathrm{QC}$ for different numbers of qubits $n_q$, is assessed as a function of the data encoding circuit repetitions (depth) $L$.``NE" represents the case where no entanglement is present in the circuit. (\textbf{b}) Summary of the performance increase as a function of the system size for $L=7$. (\textbf{c}) Entanglement capability of the data encoding circuit as a function of the depth $L$. Uncertainties are comparable to the size of the displayed data points and are computed using the same 5-fold testing procedure as in Fig.~\ref{fig:results}.}
    \label{fig:delta_qc}
\end{figure*}
We assess the performance of the proposed quantum algorithms in terms of their ability to detect new-physics events. Quantitatively, we compute the Receiver Operating Characteristic (ROC) curve and the corresponding Area Under the Curve (AUC). Even though the AUC provides an overview of the model performance, it is not an informative metric when only a region of the ROC curve is of practical interest. Such is the case for anomaly detection in \ac{HEP}. Concretely, we focus on specific values of \ac{TPR} and their corresponding \ac{FPR}. 
In \ac{HEP} nomenclature these metrics are called signal and background efficiencies and are denoted by $\varepsilon_s$ and $\varepsilon_b$, respectively. We consider typical working points in physics analyses at the \ac{LHC}, where $\varepsilon_s=0.6, 0.8$. Furthermore, to quantify differences in model performance we define the following metric, 
\begin{equation}
\Delta_{\text{QC}}(\varepsilon_s)=\frac{\varepsilon^{-1}_{\text{b}}(\varepsilon_s;Q)}{\varepsilon^{-1}_{b}(\varepsilon_s;C)},
\label{eq:delta_qc}
\end{equation}
where $\varepsilon^{-1}_{b}(\varepsilon_s;\cdot)$ is $\mathrm{FPR}^{-1}$ at a working point $\varepsilon_s$, given a specific quantum (Q) or classical (C) model. 

We systematically study the performance of our models across three parameters of interest using ideal simulations of quantum circuits. Firstly, we evaluate the performance using different new physics anomalous signatures, described in the ``Data" subsection. The second parameter of interest is the latent space dimensionality, evaluated for $\ell \in \{4,\,8,\,16\}$ for all models, and additionally $\ell =32$ for the clustering algorithms. Finally, the impact of the number of training samples is studied for values $N \in \{10, 600, 6000\}$. Our results are summarised in Fig.~\ref{fig:results}. Each quantum algorithm,  QK-means (left), QK-medians (middle), and the unsupervised kernel machine (right), respectively, is compared to the best-performing classical algorithm of similar model complexity that is trained and tested on the same data. 

The model performance on different \ac{BSM} scenarios is displayed in Fig.~\ref{fig:results}a. For this evaluation, the latent dimensionality is fixed at $\ell =8$ and the training sample contains 600 examples. We observe that there is a large variation in performance across the three signal signatures. This is expected since the studied \ac{BSM} signals differ in their similarity to \ac{SM} processes. The broad Graviton is the most similar and hence hardest to identify while the narrow Graviton is the most anomalous and thus, easier to identify in a model-independent setting. This is demonstrated by the consistent order in performance across all models and mirrored by the classical benchmarks.

Both clustering models exhibit an optimal latent space dimensionality with the best performance at a value of eight (Fig.~\ref{fig:results}b) and a drop for the largest category of $32$. The kernel-based model performs best in the latent space dimensions larger than four.
Even with a substantial compression from the original space to the smallest tested latent representation $\mathbb{R}^{300}\xrightarrow{\mathrm{AE}}\, \mathbb{R}^4$, we do not observe a dramatic drop in discrimination power for the clustering algorithms.
These results demonstrate that our \ac{AE} is capable of mapping \ac{HEP} jet data to a size that is tractable on current noisy quantum computers while preserving information used by the quantum models to discriminate anomalies. 
It is crucial to highlight that the \ac{AE} alone achieves a maximum AUC of approximately $75\%$ on the anomaly detection tasks, across the different latent space dimensions (see Methods for more information). 
Therefore, our quantum algorithms provide a substantial enhancement of performance.

Furthermore, the performance of all (Q)ML algorithms improves with increasing training size, as typically expected. The QK-medians reaches a better overall classification power compared to the classical clustering algorithm for the case of $N_\text{train}=10$ (cf. Fig.~\ref{fig:results}c), supporting the empirical observation of other studies that the performance advantage of \ac{QML} models might reside in smaller training sizes~\cite{terashi_event_2021, wu_2021_kernel, Guan_2021}.

Overall, the unsupervised kernel machine outperforms both clustering algorithms and its behaviour is further analysed in the following.

\subparagraph{\textit{Utilisation of quantum resources}.} In our numerical experiments, we observe that the unsupervised kernel machine outperforms the classical benchmark in detecting new-physics events. In Fig.~\ref{fig:delta_qc} we present a systematic investigation of the behaviour of the quantum kernel model as a function of the provided quantum resources, namely quantified by the number of qubits $n_q$ and the depth of the circuit $L$. Specifically, Fig.~\ref{fig:delta_qc}a depicts the relationship between the anomaly detection performance of the quantum model and $L$, for each $n_q$, up to 24 qubits. Here, ``NE" denotes a data encoding circuit without entanglement, achieved by removing all CNOT gates. Fig.~\ref{fig:delta_qc}b summarizes the enhancement in model performance with increasing qubit count for $L=7$. In Supplementary Fig.~4, the respective results for $\varepsilon_s=0.6$ are presented, demonstrating a similar behaviour to that observed in the $\varepsilon_s=0.8$ case. Additionally, in Fig.~\ref{fig:delta_qc}c we show the entanglement capability of the quantum kernel circuit as a function of $L$. Data encoding circuits of larger depth lead to higher entanglement capability and expressibility; for more information see the Supplementary Note 1.

We identify a regime of the employed quantum resources, particularly for $L\geq 1$ and $n_q>4$, where the unsupervised quantum kernel machine consistently outperforms the classical benchmark. For $n_q=4$, the quantum model consistently performs worse than its classical counterpart. 
We note that with the utilisation of additional quantum resources, achieved by increasing both $n_q$ and $L$, the performance of the quantum model is significantly enhanced (cf.~Fig.~\ref{fig:delta_qc}a and b). 
In particular, for each $n_q$, all models reach their maximum performance when the corresponding data encoding circuits are of high entanglement (Fig.~\ref{fig:delta_qc}c).
Furthermore, we demonstrate that in the absence of entanglement in the quantum kernel (NE), the performance of the quantum model drastically declines across all values of $n_q$. 
These observations emphasize the critical role of entanglement as a resource for the developed quantum model. 
While this holds for our \ac{HEP} dataset, previous studies on standard classical \ac{ML} benchmark datasets report that the presence of entanglement does not improve the quantum model performance~\cite{Altares-Lopez_2021, Bowles:2024fvp}.

For the classical benchmark, we choose a kernel machine equipped with the Radial Basis Function (RBF) kernel. This model consistently provided the best performance among all assessed classical models: linear, polynomial, sigmoid, and RBF kernels for kernel-based models, and the K-means and K-medians clustering algorithms presented in Fig.~3. Explicit values of $\varepsilon_b^{-1}(\varepsilon_s)$ are presented in Table~\ref{tab:var_ntrain} and Table~\ref{tab:var_latent}, for a fixed latent dimension and fixed training size, respectively. As in Fig.~\ref{fig:results}, the values and the corresponding uncertainties presented in Table~\ref{tab:var_ntrain} and Table~\ref{tab:var_latent} are obtained by evaluating the trained models on a 5-fold testing dataset of $N_\mathrm{test}=10^5$ data samples.

\subparagraph{\textit{Quantum Hardware Calculations.}}
We implement the unsupervised kernel machine on a quantum computer, namely the \verb|ibm_toronto| machine. The goal is to demonstrate the capability of the designed embedding circuit (cf. Fig.~\ref{fig:circuits}a) to run on current noisy near-term quantum devices. The size of the training and testing dataset is fixed at 100 and the number of qubits is eight. Furthermore, the number of circuit evaluations needed during inference time to construct the quantum kernel matrix is $N_\mathrm{train}\cdot N_\mathrm{test}$. Hence, the number of testing samples is constrained not by the inherent limitations of the noisy near-term quantum device, but by the number of evaluations we can submit to the device via the IBMQ cloud in a reasonable amount of time. For more details on the quantum hardware implementation see Methods.
\begingroup
\setlength{\tabcolsep}{10pt} 
\renewcommand{\arraystretch}{1.1} 
\begin{table}[t]
\centering
    \resizebox{\columnwidth}{!}{
\begin{tabular}{lccc}\toprule[1.25pt]
Working point & $N_\mathrm{train}$ & Quantum & Classical        \\ \hline
\multirow{2}{*}{$\varepsilon^{-1}_b(\varepsilon_s=0.8)$} & 6000  & $45.16 \pm 3.38$   & $22.01 \pm 1.16$ \\
               & $10$ & $\,\,\,8.49 \pm 0.23$    & $\,\,\,5.50 \pm 0.10$  \\ \midrule
\multirow{2}{*}{$\varepsilon^{-1}_b(\varepsilon_s=0.6)$}     & $6000$ & $183.71 \pm 24.40$ & $78.04 \pm 3.45$  \\
                                                       & $10$ & $27.84 \pm 0.94$   & $17.87 \pm 0.75$\\\bottomrule[1.25pt]
\end{tabular}
}
\caption{\textit{Background efficiency for different training data sizes}. The inverse of the background efficiency $\varepsilon_b^{-1}(\varepsilon_s)$ of the unsupervised quantum and classical kernel machines is presented for models trained on a fixed latent dimension of $\ell=8$, i.e., number of qubits $n_q$, corresponding to signal efficiencies of $\varepsilon_s=0.6,0.8$. The number of training samples $N_\mathrm{train}$ is varied.}
\label{tab:var_ntrain}
\end{table}
\begin{table}[t]
\centering
    \resizebox{\columnwidth}{!}{
\begin{tabular}{l c c c}\toprule[1.25pt]
Working point  & $\ell$  & Quantum & Classical        \\ \hline
\multirow{4}{*}{$\varepsilon^{-1}_b(\varepsilon_s=0.8)$} & $4$ & $\,\,\,2.42 \pm 0.02$    &\, $5.19 \pm 0.09$  \\ 
               & $8$ & $38.98 \pm 2.28$    & $11.68 \pm 0.33$  \\ 
               & $16$  & $42.14 \pm 1.92$   & $\,\,\,7.54 \pm 0.30$ \\
               & $24$  & $50.18 \pm 4.62 $   & $ \,\,\,3.42 \pm 0.08 $ \\
                
               \midrule
\multirow{4}{*}{$\varepsilon^{-1}_b(\varepsilon_s=0.6)$} & $4$ & $\,\,\,6.28 \pm 0.19$    & $26.42 \pm 1.20$  \\ 
               & $8$ & $148.07 \pm 13.52$   & $58.92 \pm 5.44$\\
               & $16$ & $122.04 \pm 14.02$ & $41.07 \pm 2.59$ \\
               & $24$  & $140.77 \pm 21.51$   & $ \,\,\,8.30 \pm 0.32$ \\                                                     
                                                       \bottomrule[1.25pt]
\end{tabular}
}
\caption{\textit{Background efficiency for different latent space dimensions}. The inverse of the background efficiency $\varepsilon_b^{-1}(\varepsilon_s)$ of the unsupervised quantum and classical kernel machines is presented for models trained on a dataset of $N_\mathrm{train}=600$ samples, corresponding to signal efficiencies of $\varepsilon_s=0.6,0.8$. The dimensionality $\ell$ of the latent space, which is equal to the number of qubits $n_q$, is varied.}
\label{tab:var_latent}
\end{table}
\endgroup

The results are presented in Table~\ref{tab:hardware}. We use the AUC to summarise the performance of the models, given the limited number of testing samples. The AUC is less prone to statistical fluctuation being the integral of the ROC curve, compared to $\varepsilon_b^{-1}$, which corresponds to a point on the ROC curve. Consequently, we note that the values obtained on hardware, due to low testing statistics, are less accurate and should not be compared to the ones obtained by the more statistically robust computation using $10^5$ testing samples (cf. Fig.~\ref{fig:results}).

The number of shots per circuit is $10^4$ and was numerically determined. Recent advancements in classical shadows techniques~\cite{Huang2020} and informationally-complete positive operator-valued measurements (IC-POVM)~\cite{PRXQuantum.2.040342} enable the evaluation of a set of generic operators, using only a logarithmic number of measurements, dramatically reducing the costs for the evaluation of the quantum kernel. Moreover, the standard deviation of the expectation values can be improved by means of a further optimisation of the IC-POVM effects~\cite{PRXQuantum.2.040342} or duals~\cite{malmi2024enhanced,fischer2024dual}. 

We find that allowing a standard deviation of order $10^{-3}$ in the kernel matrix elements, still yields the presented results where $\Delta_\mathrm{QC}>1$. Conversely, when the uncertainty reaches order of $10^{-2}$ or higher, our quantum model is not able to consistently outperform the classical benchmarks.

Additionally, the mean purity of the states $\braket{\mathrm{tr}\rho^2}\equiv \braket{k(x_i, x_i)}$, over the data points $x_i$, is measured on the quantum computer to ensure that throughout the computation the state has not decohered, i.e., lost its quantum nature due to noise and decoherence of the qubits. A fully mixed (decohered) state yields a purity of $1/2^{n_q}$, resulting in approximately $0.39\times 10^{-2}$ for eight qubits.
The hardware performance and the corresponding purities confirm that the designed quantum circuit for kernel-based anomaly detection, at least up to three repetitions, is indeed suitable for current quantum computers.

\begingroup
\setlength{\tabcolsep}{14pt} 
\renewcommand{\arraystretch}{1.1} 
\begin{table}[t]
    \centering
    \resizebox{\columnwidth}{!}{
    \begin{tabular}{lcc}\toprule[1.25pt]
    {Kernel Machine Run} & {AUC} \quad& {$\braket{\mathrm{tr}\rho^2}$}\\ \midrule
    Hardware $L=1$  $\quad$& 0.844 & 0.271(6)  \\
    Ideal $L=1$ \quad & 0.999 & 1\\\midrule
    Hardware $L=3$ \quad & 0.997 & 0.15(2)  \\
    Ideal $L=3$  \quad& 1.0  & 1\\\midrule
    Classical    \quad & 0.998  & -\\ \bottomrule[1.25pt]
    \end{tabular}
    }
    \caption{\textit{Quantum hardware results}. Comparing the performance of the unsupervised kernel machine in hardware and ideal simulation in terms of AUC, for one ($L=1$) and three ($L=3$) repetitions of the data encoding circuit. The mean purity $\braket{\mathrm{tr}\rho^2}$ is presented with its uncertainty in parenthesis at the corresponding decimal point.}
    \label{tab:hardware}
\end{table} 
\endgroup

\section{Discussion}\label{sec:discussion}
We presented a realistic study of \ac{QML} models for anomaly detection in proton collisions at the \ac{LHC}. The ability of the designed models to identify new-physics (\ac{BSM}) events was thoroughly investigated using metrics of interest in \ac{HEP} analyses. The proposed combination of an autoencoder that compresses raw \ac{HEP} jet features to a tractable size, with quantum anomaly detection models proved to be a viable strategy for data-driven searches for new physics at the \ac{LHC}. 

Previous studies have examined classification tasks both in classical \ac{ML} benchmark datasets~\cite{Altares-Lopez_2021, Bowles:2024fvp}, and in \ac{HEP} problems~\cite{Guan_2021, qmlHiggs2021, blance_quantum_2021, terashi_event_2021, wu_2021_kernel,Blance_ad_2021, alvi_quantum_2022, Ngairangbam_2022}.
These investigations generally report either a superiority of classical \ac{ML} models over quantum ones or no significant difference in performance between the two. Particularly, Refs.~\cite{Altares-Lopez_2021,Bowles:2024fvp} demonstrate that excluding entanglement from QML models does not affect the performance on conventional ML benchmark datasets. These results are in stark contrast to our findings.
In this work, we presented an instance where the developed quantum kernel model outperforms classical models for an anomaly detection task in fundamental physics. 
Furthermore, we demonstrated the importance of quantum resources, such as entanglement and the number of qubits, utilised by our data encoding circuit (Fig.~\ref{fig:delta_qc}). Additionally, we established that the designed model can be implemented on current quantum computers.
Our result is achieved using realistic datasets, and stretches beyond one-shot experiments or specific values of the considered parameters. Nevertheless, our findings do not preclude the existence of other classical algorithms that can reach similar or improved performance on our dataset. Investigating such classical approaches, such as tensor networks, constitute an interesting topic for future studies.

The data encoding circuit for the quantum kernel is designed to have favorable properties, as discussed in the Results section. However, it is not explicitly biased to features characteristic to \ac{HEP} data. Future studies need to investigate the construction of \ac{QML} models that have an inductive bias towards the structure of \ac{HEP} data to investigate for guarantees in terms of trainability and performance advantage. Moreover, a crucial objective of future research is to determine which component of the correlations between features is leveraged by the quantum model to attain an enhanced performance. Other works on \ac{QML} share also these insights in a more general context~\cite{Huang2021, Kubler_2021, Pesah_2021, Meyer_2022}. 


Data from particle physics experiments still exhibits quantum effects and necessitates a quantum field theory framework for accurate description despite undergoing classical processing, compression, and storage. Such quantum effects include spin correlations~\cite{top_correlations_CMS2019}, entanglement~\cite{ATLAS:2023_entanglement, Cervera2017, Severi2022, Fabbrichesi2023}, and Bell inequality violation~\cite{Fabbrichesi2021, Afik_Nova_2022, Ghosh2023}. In our work, we show that entanglement plays a crucial role in the ability of the quantum kernel machine to identify anomalies. Previous studies have defined metrics to quantify the suitability of specifically engineered classical datasets for quantum models constructed with the appropriate inductive biases~\cite{Kubler_2021, Huang2021}. Yet, a widely applicable measure to evaluate the degree of ``quantumness" and appropriateness of a classical dataset for quantum machine learning remains still elusive, highlighting a critical area for future investigation.

Recent studies stress the importance of investigating datasets that are generated by fundamentally quantum processes~\cite{Kubler_2021, Gyurik:2023quj, Huang2021}. The \ac{HEP} dataset we investigate falls in this category. We hope that our work will motivate both the \ac{HEP} and \ac{QML} communities, fostering further investigations and scientific discussion about the potential benefits of using quantum approaches, based on appropriate benchmark studies using datasets and methodologies also inspired by this work.  

Additionally, we suggest that it is important to expose the quantum algorithms to features of minimal post-processing avoiding manual extraction of classical high-level features. In fact, the results of similar works reinforce this claim~\cite{Belis:2024ctj, Schuhmacher2023}, particularly in Ref.~\cite{Schuhmacher2023} since the considered \ac{HEP} dataset is described by high-level features.

We anticipate that our results, along with the growing literature in \ac{QML} for physics data and the parallel rapid improvement of available quantum hardware, can stimulate fruitful research at the intersection of hybrid quantum-classical algorithm design and fundamental physics data processing. Advancements along these lines can potentially lead to novel architectures and quantum-classical synergies, to performance advantages in well-defined sets of problems, and to the enhancement of our fundamental understanding of both fields. 

\section*{Methods}
\subparagraph{\textit{\textbf{Data preparation}}.}Events are produced with \PYTHIA~\cite{pythia} and are processed using the particle-flow algorithm~\cite{delphes}. For all physical quantities, we use natural units, i.e., $c=\hbar=1$. The width of the \ac{RS} graviton is set to a negligible value, fixing the $\kappa \times m_G$ parameter~\cite{Bijnens:2001gh} in \PYTHIA to 0.01. The graviton width-to-mass ratio is fixed by choosing $\kappa \times m_G = 2.5$, which results in a width-to-mean ratio $\sim 35\%$ for $\mjj$ after detector effects. The resonance masses are varied between 1500\GeV and 4500\GeV, in steps of 1000\GeV.

A jet training dataset is defined using the so-called dijet $\Delta \eta$ sideband, requiring the pseudorapidity separation between the two highest-$\pt$ jets in the event $|\Delta \eta_{jj}|>1.4$. This kinematic region is largely populated by QCD multijet events and any potential signal contamination by BSM processes involving high-energy collisions is typically negligible. Our approach is loosely inspired by Ref.~\cite{cms2020_dijet}, and we assume that \ac{BSM} particles are produced via s-channel processes. The classical autoencoder is trained to compress particle jet objects individually, using background QCD data in the sideband region, without access to truth labels. 
The quantum and classical \ac{AD} algorithms are trained on the latent representation of QCD multijet events in a signal region, defined requiring $|\Delta \eta_{jj}|\leq1.4$. After training, their performance is assessed using background and anomalous events belonging to the same signal region.

Jets are clustered from reconstructed particles using the anti-$\kt$~\cite{antikt} clustering algorithm with a jet-size parameter $R=0.8$ in \FASTJET~\cite{fastjet}. 
Dijet events are selected requiring two jets with $\pt>200\GeV$ and within $|\eta|<2.4$. 
In order to emulate the effect of a typical \ac{LHC} online event selection, only jet pairs with invariant mass $\mjj>1260\GeV$ are considered. 

Each event is represented by its two highest-$\pt$ jets. Each jet is represented by a list of constituents, built considering the 100 highest-$\pt$ particles,
reconstructed by the particle-flow algorithm and contained in a cone of radius $\Delta R = \sqrt{\Delta \phi^2 + \Delta \eta^2}<0.8$ from the jet axis. Particles are ordered according to their decreasing $\pt$ value. Zero padding is used to extend particle sequences shorter than 100 particles. Each particle is represented by its $\pt$ and its $\eta$ and $\phi$ distance from the jet axis. In its {\it raw} format, a jet is then represented as a $100\times 3$ matrix. 

The data samples are processed with {\tt DELPHES} library~\cite{delphes} to emulate detector effects, using the distributed CMS detector description. The effect of additional simultaneous proton collisions (\textit{in-time pileup}) is included. This is achieved by overlapping low-momentum collision events, sampled from a set of simulated \ac{LHC} collisions, with the primary interaction vertex.
The number of overlayed events varies according to a Poisson distribution centered at 40, resembling the in-time pileup recorded at the \ac{LHC} at the end of 2018. The QCD multijet dataset size corresponds to an integrated luminosity of $64\fbinv$, comparable to running the \ac{LHC} for about one year.

\subparagraph{\textit{\textbf{Autoencoder.}}}To reduce the dimensionality of our low-level feature inputs, we developed an autoencoder, whose architecture is displayed in Supplementary Fig.~2. The encoder consists of 1D and 2D convolutional layers, followed by pooling and dense layers. The decoder mirrors the architecture of the encoder. The bottleneck width defines the dimensionality of the latent space representation. Weights are He-uniform initialised and all layers feature ELU activations, except for a $\tanh$ activation right before the bottleneck, that constrains the latent space to a range $z \in (-1,1)$. The model is trained for 200 epochs with learning rate decay, minimizing the Chamfer-loss~\cite{fan2016point} defined as a pairwise reconstruction distance 
\begin{equation}\label{eq:ae_loss}
    L_R = \sum_{i \in \text{input}} \min_{j}\left[(x^{(i)} - x^{(j)})^2\right] + \sum_{j \in \text{output}} \min_{i}\left[( x^{(j)} - x^{(i)})^2\right].
\end{equation}
Inputs are standard-normalised at the entry to the encoder and outputs are de-normalised in the last layer of the decoder. 
The designed autoencoder achieves the desired balance between model complexity, effective compression of \ac{HEP} data, and architectural stability. 
To verify the stability of the model, various initialisations of the training have been tested using the same dataset. 

The \ac{AE} model can also be used in standalone anomaly detection mode by using the reconstruction loss as a metric to identify anomalies: events with a reconstruction loss higher than a chosen threshold are flagged as anomalies. 
As for all studied models, we vary the threshold on the reconstruction loss to quantify the performance of the autoencoder by constructing the ROC curves and computing their corresponding AUC. The \ac{AE}, as presented in the ``Evaluation of model performance" subsection in the Results, achieves an AUC of approximately $75\%$ consistently for all tested latent dimensions. 
This additional classical benchmark demonstrates the improved anomaly detection performance that our proposed paradigm provides.

\subparagraph{\textit{\textbf{Quantum Kernel machines.}}} Given the quantum feature map $\rho:\; \mathcal{I}\mapsto \mathcal{H}$, that maps the input \ac{HEP} event space $\mathcal{I}$ to the Hilbert space $\mathcal{H}$ of $n_q$ qubits, we can define the quantum kernel using the Hilbert-Schmidt inner product
\begin{equation}
k(x_i,x_j)\coloneqq \text{tr}[\rho(x_i)\rho(x_j)] 
= 
\left|{\braket{0|U^\dagger(x_i) U(x_j)|0}}\right|^2,
    \label{eq:quantum_kernel}
\end{equation}
where the quantum feature map is identified as the density matrix operator $\rho(x_i)\coloneqq U(x_i)\ket{0}\bra{0}U^\dagger(x_i),$ where $U(\cdot)$ corresponds to a data encoding quantum circuit (Fig.~\ref{fig:circuits}a) that represents the quantum feature map, and $\ket{0}=\ket{0}^{\otimes n}\in\mathcal{H}$. The unsupervised kernel machine is trained to find the hyperplane that maximizes the distance of the training data from the origin of the feature vector space, similar to the classical algorithm~\cite{one_class_svm}. We optimize the objective function on a classical device, while the kernel matrix elements $K_{ij}=k(x_i,x_j)$ are sampled from a quantum computer or a quantum simulation on a classical device. The data encoding circuit, presented in Fig.~\ref{fig:circuits}a, is heuristically designed, guided by theoretical results regarding the increase of model expressivity as a function of the repetitions (depth) of the ansatz~\cite{sim_expressibility_2019, Schuld_Sweke_Meyer_2021}, and the data re-uploading strategy which is known to construct powerful quantum classifiers~\cite{Perez2020, Jerbi2023}.

\subparagraph{\textit{Training the unsupervised kernel machine.}} We present the optimisation task of the unsupervised kernel machine in its primal form adopting the notation from Ref.~\cite{one_class_svm}  for completeness. Given a set of training data $x_1, x_2,\dots, x_\mathrm{N}\in\mathcal{X}$, where $\mathcal{X}$ is an arbitrary vector space, N is the number of training data points, and $x_i$ represent the feature vectors of the data, one can write the objective function of the model and its corresponding constraints as
\begin{align}
    &\min_{w\in\mathcal{F},\,\xi\in\mathbb{R}^\mathrm{N},\,\rho\in\mathbb{R}}\quad \frac{1}{2}||w||^2+\frac{1}{\nu\mathrm{N}}\sum_i\xi_i -\rho\label{eq:opt1}\\
    &\mathrm{subject\; to}\quad w\cdot\Phi(x_i)\geq \rho-\xi_i,\,\xi_i\geq 0,\; \forall i,\label{eq:opt2}
\end{align}
where $\Phi:\mathcal{X}\mapsto\mathcal{F}$ is the feature map, which is used to define the kernel $k(x_i,x_j)=\Phi(x_i)\circ\Phi(x_j)$, $\mathcal{F}$ is the feature space, $w$ are the trainable weights of the model, $\xi_i$ are the slack variables, $\rho$ is the distance of the hyperplane from the origin of the feature space, and $\nu\in(0,1)$ is a hyperparameter. Training the kernel machines corresponds to solving the optimisation problem above with respect to $w$ and $\rho$.  It can be shown that $\nu$ is an upper bound on the fraction of outliers, or anomalies following the nomenclature of this manuscript, in the training dataset~\cite{one_class_svm}. 

An optimal $\nu$ cannot be identified globally for every possible signal distribution, i.e., \ac{BSM} scenario, in the unsupervised anomaly detection context. On the contrary, in a supervised learning setting where the specific signal and background are known by construction, one can hyperoptimise the \ac{ML} algorithm using model selection techniques. Choosing a value of $\nu$ based on hyperoptimisation with respect to a specific \ac{BSM} dataset can lead to biasing the model towards this particular anomaly distribution. Although this can increase the performance of the model on the chosen \ac{BSM} signature, the sensitivity of the anomaly detection strategy can be diminished for other new-physics scenarios. Hence, such a procedure can lead to overall reduced performance of the anomaly detection model and is not consistent with the initial motivation behind model-independent searches of new physics at the \ac{LHC} experiments as presented in the Introduction.

For our study, we choose $\nu=0.01$ and keep it fixed for all investigated models and \ac{BSM} datasets. Therefore, we assume that the model should, falsely, flag at most $1\%$ of the QCD background samples used for training. Furthermore, given the small training sizes used for the quantum models, e.g. 600, smaller values of $\nu$ would lead to a none-integer upper bound on the number of outliers causing the optimisation problem in Eq.~\ref{eq:opt1} and~\ref{eq:opt2} to be ill-defined.

\subparagraph{\textit{\textbf{QK-means.}}} Given two input feature vectors \textit{u} and \textit{v}, the quantum states $\ket{\psi}$ and $\ket{\phi}$ are prepared as
\begin{equation}
\ket{\psi} = \frac{1}{\sqrt{2}}(\ket{0}\ket{u} + \ket{1}\ket{v}), \quad \ket{\phi} = \frac{1}{\sqrt{Z}}(|u|\ket{0} - |v|\ket{1}),
\end{equation}
where Z = $|u|^2 + |v|^2$.
Subsequently, a controlled swap test (Fig.~\ref{fig:circuits}b) is performed to calculate the overlap $\braket{\phi|\psi}$. Then, the distance $D$, proportional to the Euclidean distance is calculated as
\begin{equation}
D = 2 Z |\braket{\phi|\psi}|^2.
\end{equation}
To find the closest cluster, the minimisation technique starts by selecting a random threshold picked from the vector of distances and constructing a threshold-oracle (Eq.~\ref{eq:oracle}) as a linear combination of \textit{m} simpler oracles, where $m \leq k$. A simple oracle is a unitary operator performing $\ket{x}\ket{q} \congto \ket{x}\ket{q \oplus f(x)}$ where $\ket{x}$ is an array of indices, $\oplus$ is the addition modulo 2 and $\ket{q}$ is the oracle control qubit. The threshold oracle 
\begin{equation}
O = \delta(t_0 - 1)O_0 + \delta(t_1 - 1)O_1 + \cdots + \delta(t_m -1)O_m 
\label{eq:oracle}
\end{equation}
marks multiple entries by linking a set of simple oracles through the $\delta$ function centered at threshold $t_i$ where oracle $O_i$ marks all inputs $\{x\}$ for which $f(x) < t_i$ and where $f$ returns the value of the distance array at index $x$.
The minimisation procedure runs for $\sqrt{2^{|k|}}$ iterations, applying the Grover search algorithm, with time complexity of $\mathcal{O}(\sqrt{k})$,  which sets the most probable output as the new threshold for the next iteration. As the last step, we calculate the mean of the centers classically.

\subparagraph{\textit{\textbf{QK-medians.}}} The quantum circuit in Ref.~\cite{qkmeans_noisy} uses destructive interference probabilities to find a distance proportional to the Euclidean distance between N-dimensional points. To prepare a quantum state $|\psi\rangle$, 

\begin{equation}
|\psi\rangle = [a_1^\prime \ldots a_n^\prime \; b_1^\prime \ldots b_n^\prime]^T,
\label{eq:psi_qkmedians}
\end{equation}
where the classical input vectors, $(a_1, \ldots, a_n)$ and $(b_1, \ldots, b_n)$, are normalised and padded with zeros if $log_2(n) \not\in \mathbb{Z}$,
\begin{equation}
a_i^\prime = \frac{a_i}{\mathcal{N}}; \; b_i^\prime = \frac{b_i}{\mathcal{N}},
\label{eq:norm_inputs}
\end{equation}
and the normalisation $\mathcal{N}$ is given by
\begin{equation}
\mathcal{N} = \sqrt{a_1^2 \,+\, \ldots \,+\,a_n^2 \,+\, b_1^2 \,+\, \ldots \,+\,b_n^2}.
\label{eq:norm}
\end{equation}
Then, one Hadamard gate is applied to the most significant qubit (MSQB) and destructive interference probabilities are obtained by measuring the MSQB in state \textbar1$\rangle$. This circuit only prepares one quantum state and applies only one quantum gate, making it less complex than the circuit for computing distances in the QK-means algorithm. 

After the distance calculation, all points are assigned to a specific cluster using a hybrid minimisation procedure. The median, being the point with the minimal distance to all other points in the cluster, is found iteratively by searching in the direction of more concentrated points~\cite{l1_median_real}. For both clustering models the number of clusters, $k$, is set to two and the iterative procedure is repeated until convergence is reached up to a tolerance of $\varepsilon$.

\subparagraph{\textit{Quantum hardware implementation.}} The unsupervised quantum kernel machine has been implemented for both the quantum hardware and simulation computations using \verb|qiskit|~\cite{qiskit}.  The data encoding circuit (see Fig.~\ref{fig:circuits}) used to construct the feature map of the quantum kernel needs to be transpiled to the gates that are native to the hardware device: $R_z, \sqrt{X}, X, $ and $CX$(CNOT). Due to the presence of noise in quantum hardware, the sampled kernel matrix need not be positive semi-definite (PSD). 
For this reason, the matrix that we retrieve from the hardware is transformed to the closest PSD matrix to ensure that the convex optimisation task of the algorithm is well-defined. Furthermore, the qubits of the algorithm need to be mapped to the physical qubits of the hardware, which for the \verb|ibm_toronto| machine follow the topology depicted in Supplementary Fig.~3. 

We execute our models using the following physical qubits: $[5, 8, 11, 14, 16, 19, 22, 25]$, corresponding to the indexing in Supplementary Fig.~3. This choice is based on the calibration data provided at the time of submitting the run, taking into account the estimated gate and measurement noise. Simultaneously, we ensure nearest-neighbor connections between the physical qubits to not introduce unnecessary SWAP gates which significantly increase the depth of the transpiled circuit. 

\section*{Author Contributions}

V.B. designed the data encoding quantum circuit for the kernel machines, implemented and trained the corresponding quantum and classical models, related the observed performance advantage to the studied properties of the circuit, and implemented experiments on a quantum computer. 
M.P. and K.A.W. prepared the dataset used in this work. K.A.W. designed and trained the classical autoencoder to compress the data, designed and trained the QK-means algorithm and trained its classical equivalent. 
E.P. designed and trained the QK-medians algorithm and trained its classical equivalent. 
All authors, V.B., K.A.W, E.P., P.B., G.D., M.G., M.P., F.R., I.T., S.V., contributed to defining the research strategy, reviewing the results, and preparing this manuscript.

\section*{Data availability}\label{sec:data}
The datasets generated for this study are publicly available on Zenodo~\cite{zenodo_data}.

\section*{Code availability}\label{sec:code}
The code developed for this paper is available publicly in the GitHub repository: \url{https://github.com/vbelis/latent-ad-qml}. All source code is also archived on Zenodo~\cite{qad23}. The Qiskit~\cite{qiskit} framework is used to implement the quantum circuits of the unsupervised kernel machine and the QK-means algorithm, and Qibo~\cite{qibo} framework is used for QK-medians. For the expressibility and entanglement capability computation we used a software implementation that we have released publicly~\cite{muser_belis_2022}.

\section*{Acknowledgements}
V.B. is supported by an ETH Research Grant (grant no. ETH C-04 21-2). V.B. would like to thank Adrián Pérez Salinas and Elias Zapusek  for the valuable discussions and insights.
E.P., K.A.W., and M.P. are supported by the European Research Council (ERC) under the European Union's Horizon 2020 research and innovation program (grant agreement n$^o$ 772369).
M.G. and S.V. are supported by CERN through the Quantum Technology Initiative. F.R. acknowledges financial support by the Swiss National Science Foundation (Ambizione grant no. PZ00P2$\_$186040).
Access to the IBM Quantum Services was obtained through the IBM Quantum Hub at CERN. The views expressed are those of the authors and do not necessarily reflect the official policy or position of IBM or the IBM Quantum team.

\section*{Competing Interests}
The authors declare no competing interests.

\bibliography{main}

\onecolumngrid
\renewcommand\thesection{\large Supplementary Note~\arabic{section}}
\renewcommand{\figurename}{Supplementary Fig.}
\setcounter{figure}{0}
\setcounter{section}{0}
\section{Expressibility, entanglement capability, and data dependence}\label{ap:expr_ent}
\subparagraph{\textit{Background}.} Expressibility and entanglement capability are measures, proposed in Ref.~\cite{sim_expressibility_2019}, to quantitatively describe properties of Parametrised Quantum Circuits (PQC). Expressibility measures the ability of a given PQC of $n$ qubits to explore the corresponding Hilbert space.  A PQC of high expressibility is expected to generate, by sampling its parameters uniformly, (pure) quantum states that uniformly cover the Hilbert space. Specifically, expressibility is defined as the difference between the PQC-generated distribution of states and that of the ensemble of Haar-random states. The difference between the two distributions is quantified by the Kullback–Leibler divergence,
\begin{equation}
\mathcal{E} = D_\mathrm{KL}\left(P_\mathrm{PQC}(F; \theta)\, || \, P_\mathrm{Haar} (F)\right),
\label{eq:expr}
\end{equation}
where $\theta$ represents the parameters of the circuit which are sampled from the uniform distribution over $[0,2\pi]$, $F$ is the fidelity $F=|\braket{\psi_\theta|\psi_\phi}|^2$, between two PQC-generated states, $P_\mathrm{PQC}(F; \theta)$ is the probability density function of fidelities obtained by the states generated by the PQC, and 
\begin{equation}
    P_\mathrm{Haar}(F)= (N-1)(1-F)^{N-2},
\end{equation}
is the probability distribution of fidelities of the Haar random states, where $N=2^n$ the dimension of the Hilbert space. By discretizing the distributions $P_\mathrm{Haar}$ and $P_\mathrm{PQC}$ using histograms, we can numerically calculate the expressibility of a quantum circuit. An expressive quantum circuit will obtain expressibility scores that approach zero as seen in Eq.~\ref{eq:expr}.

The entanglement capability of a circuit is a measure of entanglement between the qubits based on the Meyer-Wallach measure $Q$~\cite{sim_expressibility_2019,Meyer_2002}. We can estimate the measure numerically via sampling the parameters of the circuit uniformly in $[0,2\pi]$ and computing the mean,
\begin{equation}
    \braket{Q}=\frac{1}{|\Theta|}\sum_{\theta\in\Theta}Q(\ket{\psi_\theta}),
    \label{eq:ent}
\end{equation}
where $\Theta$ is the set of parameters, correspondingly $|\Theta|$ the number of parameters, and $Q$ is the Meyer-Wallach measure. For more details and insights regarding these definitions refer to Ref.~\cite{sim_expressibility_2019}.

\subparagraph{\textit{Data dependence.}} The measures above are defined within the context of PQC-based algorithms, such as the variational quantum eigensolver. When dealing with data, as in this work, the notion of expressibility and entanglement capability needs to include the data distribution at hand. We extend the definitions in Eq.~\ref{eq:expr} and~\ref{eq:ent}, similarly to recent work in Ref~\cite{Thanasilp_2022}, to a data-dependent setting. Specifically, we sample the quantum circuit parameters from the distributions of the data features instead of the uniform distribution over $[0,2\pi]$. The input of the quantum models is the latent representation of the \ac{HEP} datasets, described by the latent feature vector $z$ obtained by the encoder network (see corresponding the Methods section in the main manuscript). Depending on the data input, $z$ is distributed according to the QCD \textit{background} distribution (SM) or the \textit{signal}, anomalous (BSM) probability distribution. Hence, the expressibility measure can be written as,
\begin{equation}
     \mathcal{E}=D_\mathrm{KL}\left(P_{U(z)}(F; z)\, || \, P_\mathrm{Haar} (F)\right),
\label{eq:expr_data}
\end{equation}
where $P_{U(z)}(F; z)$ represents the fidelity distribution of the states generated by the data encoding ansatz $U$ (see Fig.~2a in the main manuscript) when the features are sampled from an arbitrary probability density function $\mathcal{P}(z)$. Similarly, for the entanglement capability we can calculate $\braket{Q}$ in Eq.~\ref{eq:ent} by sampling from $\mathcal{P}(z)$.

\begin{figure*}[t]
    \centering
    \hspace*{-0.5cm}
    \includegraphics[width=\linewidth]{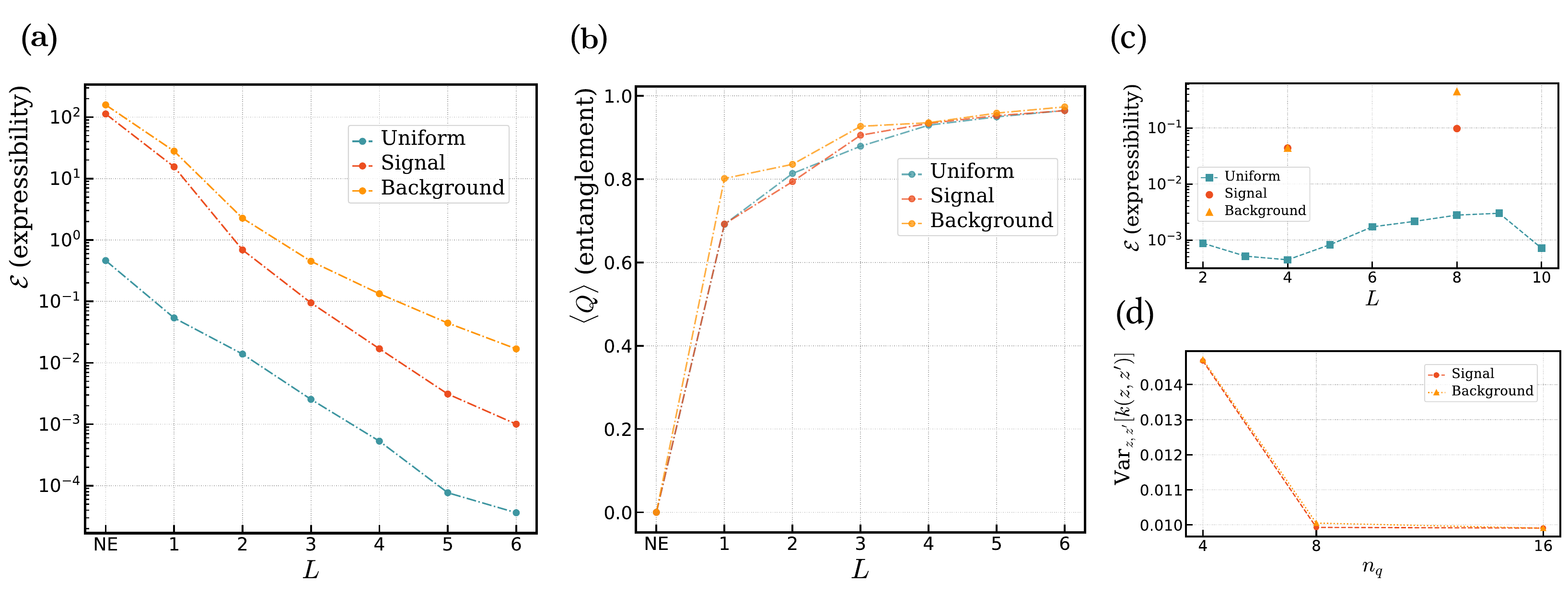}
    \caption{\textit{Characterization metrics of the data encoding circuit.} The metrics are calculated via sampling the circuit parameters from three different distributions as depicted in the legends: the uniform distribution in $[0,2\pi]$, the QCD background data distribution, and the signal (anomaly) scalar boson data distribution. (\textbf{a}) The expressibility (Expr) as a function of the different circuit architectures. (\textbf{b}) The entanglement capability $\langle \mathrm{Q} \rangle$ of the data encoding circuit as a function of the different circuit architectures. (\textbf{c}) The expressibility of the data encoding circuit as a function of the number of qubits ($n_q$). (\textbf{d}) The variance of the kernel $\mathrm{Var}_{z, z'}k(z,z')$ as a function of the number of qubits, where $k(z,z')$ is the kernel corresponding to the data encoding circuit , $z$ and $z'$ are data feature vectors sampled from the signal or background distributions.}
    \label{fig:expr_ent_var}
\end{figure*}

We compute the expressibility and entanglement capability of the designed circuit, for the case of eight qubits, as a function of the architecture variations presented in Results and in Fig.~4. For this computation, we use the software implementation in Ref.~\cite{muser_belis_2022}. The results are presented in Fig.~\ref{fig:expr_ent_var}a and~ \ref{fig:expr_ent_var}b, where the three different curves correspond to three different choices of $\mathcal{P}(z)$: sampling from the uniform distribution in $[0, 2\pi]$, the QCD background, and the signal distribution of the scalar boson \ac{BSM} scenario, respectively. The number of samples is $6\times 10^5$ for each data-dependent computation in Fig.~\ref{fig:expr_ent_var}. We verify that by increasing the depth of the proposed data encoding circuit its expressibility and entanglement capability are increased for all assessed data distributions. These results combined with Fig.~4 demonstrate a correlation between the performance of the unsupervised kernel machine on \ac{HEP} data and quantum properties of the designed feature map.

\section{Trainability of the kernel machine and exponential concentration}\label{ap:trainability}
One of the advantages of kernel-based (Q)ML is the theoretical guarantee of a global extremum of the objective function that can be found through convex optimisation~\cite{schuld_qml_is_kernel2021}. Recently, in Ref.~\cite{Thanasilp_2022}, the trainability of quantum kernel methods in \ac{QML} is studied and challenged within the scope of \textit{exponential concentration}, i.e. the case where all elements of the kernel matrix converge exponentially fast to a single value and render the model ineffective.  Therein, different sources of exponential concentration are identified, such as the expressibility of the circuit and the use of global measurements.

In our study, we do not observe such limitations based on the performance assessment of the unsupervised quantum kernel machines (cf. Results). Here, we investigate further the scaling of the expressibility of the data encoding circuit and the variance of the corresponding quantum kernel with the number of qubits. These computations provide insights on the trainability of our models beyond the tested datasets. In Fig.~\ref{fig:expr_ent_var}c and Fig.~\ref{fig:expr_ent_var}d we present the expressibility and variance of the kernel, respectively. The available $6\times 10^5$ data samples were sufficient to evaluate accurately the expressibility for four and eight qubits, but not for $n_q\geq16$. Hence, only results for four and eight qubits are presented in Fig.~\ref{fig:expr_ent_var}c when sampling the circuit parameters from the signal and background data distributions. 

Overall, an almost constant dependence between the expressibility and the number of qubits is observed in Fig.~\ref{fig:expr_ent_var}c, contrary to one of the assumptions of Theorem 1 in Ref.~\cite{Thanasilp_2022}, where the expressibility of the data encoding circuit be exponentially close to a 2-design. That is, $\mathcal{E}\in\mathcal{O}(1/b^n)$, where $b>1$ and $n$ is the number of qubits. In the case of kernel variance scaling, no exponential decay seems to be manifest.  Thus, our calculations suggest that the proposed unsupervised kernel machine does not suffer from exponential concentration due to expressibility or global measurements. Nevertheless, our numerical experiments are restricted to the presented number of qubits. \ac{HEP} datasets can be generated specifically with the purpose of assessing quantum kernel properties more thoroughly, which can be the subject of future studies.

\begin{figure*}[thb]
    \centering
    \hspace*{-0.5cm}
    \includegraphics[width=1.05\textwidth]{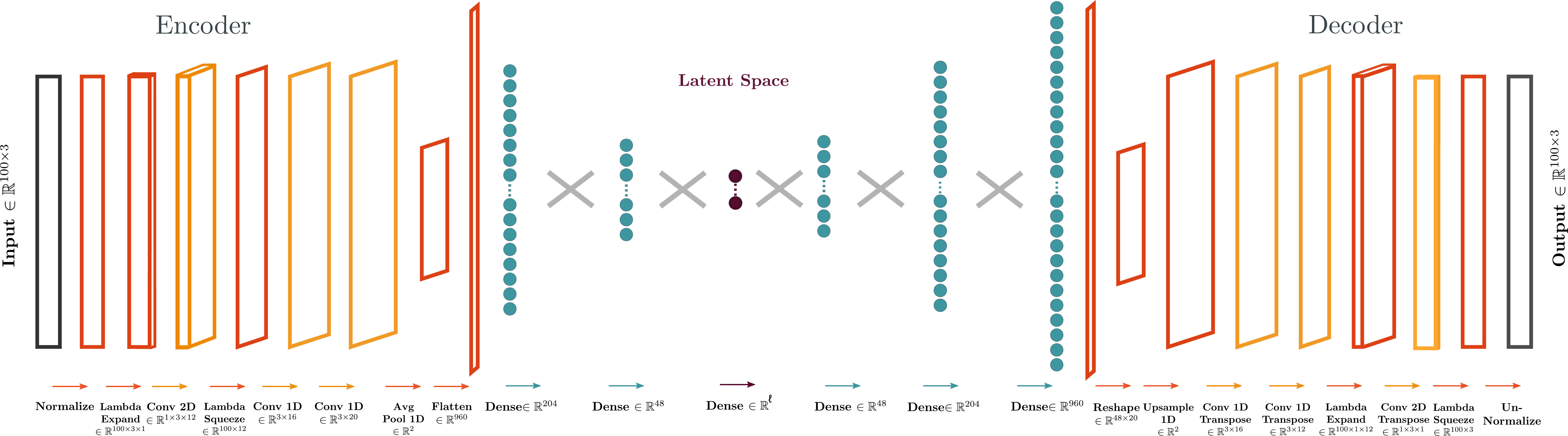}
    \caption{\textit{The Autoencoder architecture.} The model reduces the dimensionality of the high energy physics dataset from 300 dimensions per jet to latent space dimension $\ell$. The generated latent space serves as the input to the anomaly detection algorithms.}
    \label{fig:ae_arch}
\end{figure*}
\begin{figure*}[thb]
    \centering
    \includegraphics[width=0.8\linewidth]{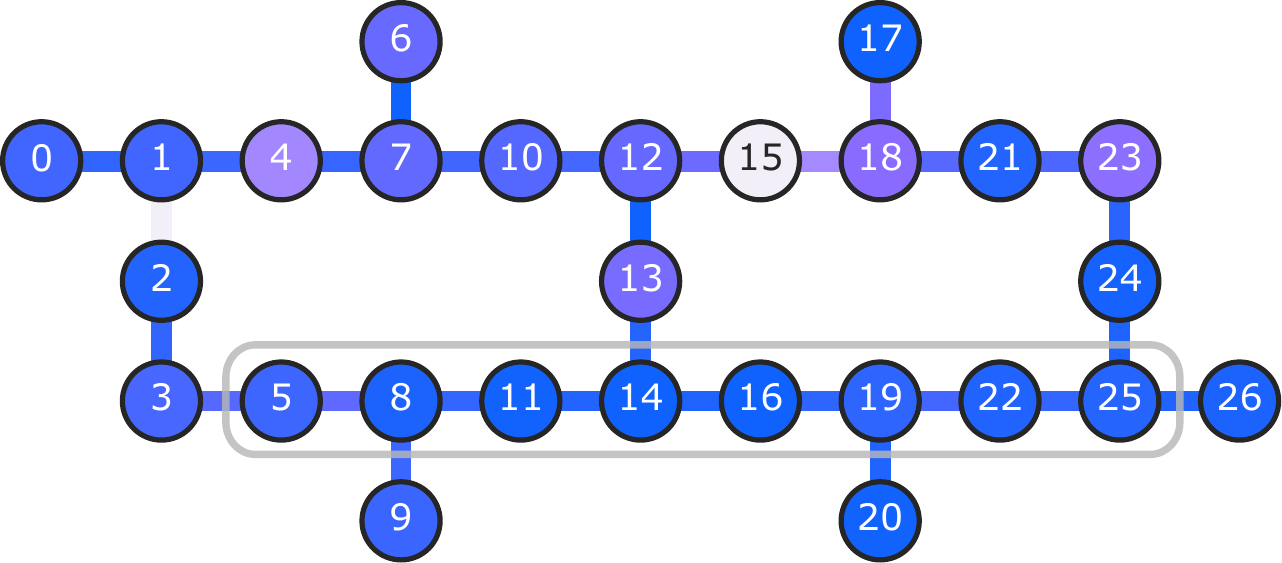}
    \caption{\textit{Topology of physical qubits}. The layout of the physical qubits on the \texttt{ibm\_toronto} machine. The connections between nodes represent the possibility of executing 2-qubit gates between neighboring qubits. The selected eight qubits, circled in grey, represent the ones used to run the quantum kernel machine. Different noise levels are color-coded, the lighter color representing higher noise levels, for single-qubit and 2-qubit gates on the nodes and connections, respectively.}
    \label{fig:hardware_qubits}
\end{figure*}
\begin{figure*}[thb]
    \centering
    \includegraphics[width=0.8\linewidth]{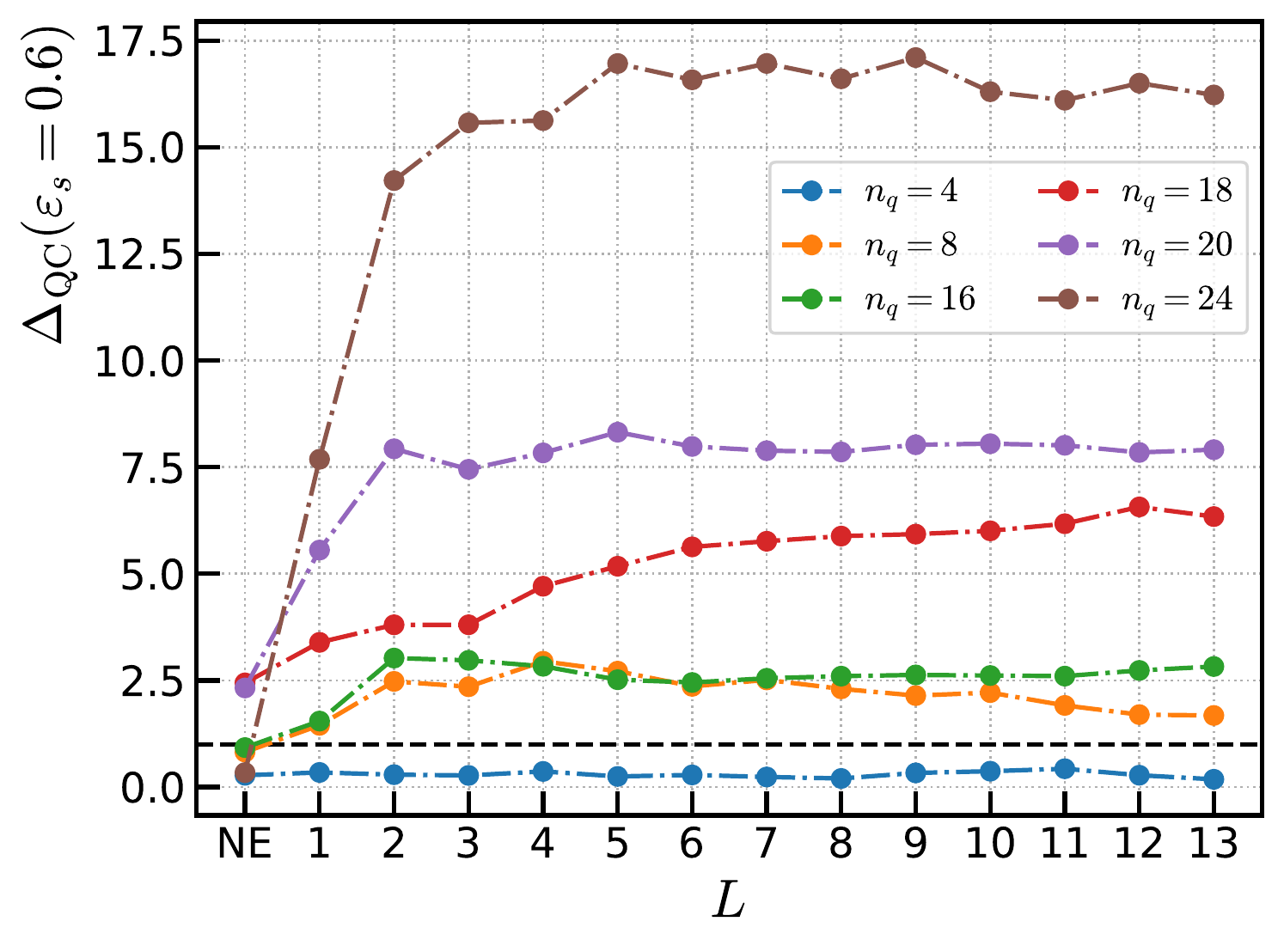}
    \caption{\textit{Performance of the unsupervised quantum kernel machine and role of entanglement} The performance of the unsupervised quantum kernel machine, quantified by $\Delta_\mathrm{QC}$, for $\varepsilon_s=0.6$ and for different numbers of qubits $n_q$, is assessed as a function of the data encoding circuit repetitions (depth) $L$. ``NE" represents the case where no entanglement is present in the circuit.}
    \label{fig:delta_qc_e_s0.6}
\end{figure*}


\end{document}